\documentclass[twocolumn,english,aps,prx,reprint,longbibliography]{revtex4-2}
\usepackage[T1]{fontenc}
\usepackage[latin9]{inputenc}
\usepackage{array}
\usepackage{float}
\usepackage{textcomp}
\usepackage{amsmath}
\usepackage{graphicx}

\makeatletter

\providecommand{\tabularnewline}{\\}

\usepackage{url}
\usepackage{comment}
\usepackage{hyperref}
\usepackage{etoolbox}

\makeatother

\usepackage{babel}
\begin{document}
\title{Nonlinear spectral back transfer limits the temporal coherency of
zonal modes}
\author{Rameswar Singh$^{1}$ }
\email{rsingh@ucsd.edu}

\author{P H Diamond$^{1,2}$}
\email{pdiamond@ucsd.edu}

\address{$^{1}$Department of Astronomy and Astrophysics, University of California
San Diego, 9500 Gilman Dr, La Jolla, CA 92093, United States of America}
\address{$^{2}$Department of Physics, University of California San Diego,
9500 Gilman Dr, La Jolla, CA 92093, United States of America}
\begin{abstract}
Zonal modes are central to magnetic confinement because their radial
shears regulate turbulence and transport. While the generation of
these flows is well understood, the mechanisms limiting their persistence
in collisionless regimes remain unresolved. In this paper, we demonstrate
that nonlinear spectral back-transfer of free energy from zonal modes
to turbulence is a major nonlinear process that limits the temporal
coherency of the shearing field. Back-transfer events induce stochastic
phase and amplitude scattering of zonal shear that limits its auto-coherence
time. Using local gyrokinetic GENE simulations for collisionless electrostatic
ion temperature gradient (ITG) driven turbulence with adiabatic electrons
for parameters representative of core conditions, we show that back-transfer
is highly intermittent and occurs in bursts that co-exist with the
zonal flow generation process. The probability distribution of the
zonal free energy transfer is non-Gaussian, with positive triangularity
(PT) exhibiting substantially higher kurtosis than negative triangularity
(NT), reflecting the markedly more intermittent and heavy-tailed character
of back-transfer bursts in PT. We find that NT plasmas exhibit significantly
reduced back-transfer compared to PT. This suppression increases the
shear auto-coherence time $\tau_{E}$ and the shearing Kubo number
$K_{u}$, leading to more resilient and effective turbulence regulation
despite lower absolute zonal kinetic energy. These results identify
back-transfer as a key nonlinear damping mechanism and suggest that
it must be explicitly treated in reduced models of drift-wave zonal-flow
turbulence. 
\end{abstract}
\maketitle

\section{Introduction \label{sec:int}}

Zonal flows are ubiquitous in nature and appears in astrophysical
disks\citep{Johansen_2009,Dittrich_2013,Bai_2014}, solar convection
zone\citep{Howe_2000}, planetary environments\citep{Vasavada_2005,Vallis2017},
earth's core\citep{Miyagoshi2010}, oceans\citep{Maximenko_2005},
laboratory fluid\citep{Busse_1994} and magnetized plasma systems.
Across all these settings, a recurring and fundamental question concerns
not merely the existence of zonal flows, but the coherence and persistence
of their shearing action --- for it is the lifetime of the shearing
field, as much as its amplitude, that determines its ability to regulate
transport. A particularly consequential example is the stratospheric
polar vortex, whose edge acts as a mixing barrier: its coherence time
directly governs the depth and duration of the Antarctic ozone hole
by controlling the isolation of polar air from ozone-rich mid-latitudes,
with coherence-breaking sudden stratospheric warming events eroding
the vortex and closing the hole\citep{JuckesMcIntyre1987,McIntyrePalmer1983,Baldwin2001}.
In the oceans, alternating zonal jets observed in satellite altimetry\citep{Maximenko_2005}
exhibit intermittently disrupted coherence whose variability controls
cross-jet mixing of heat, oxygen, and chemical tracers, with direct
consequences for deep-ocean ventilation and the global overturning
circulation. In the Earth's liquid outer core, intermittent loss of
zonal flow coherence plausibly linked to geomagnetic excursions and
reversals\citep{Finlay2003}. In astrophysical contexts, zonal flows
in protoplanetary and accretion disks modulate angular momentum transport
driven by magnetorotational turbulence, and their coherence time directly
affects the formation of pressure bumps that seed planetesimal growth\citep{Johansen_2009,Dittrich_2013,Bai_2014}.
Thus, across all these systems --- ranging from stratospheric chemistry
to astrophysical disk evolution --- the coherence time of the zonal
flow shearing field emerges as the key quantity governing transport
regulation, yet what sets this coherence time in the saturated state
remains an open and fundamental question. 

In tokamaks, zonal flows---the azimuthally symmetric, radially varying
$E\times B$ flows--- self-organize in drift-wave turbulence\citep{diamond:05,fujisawa:09}.
By shearing turbulent eddies\citep{BDT:1990,diamond:05}, they regulate
fluctuation amplitudes, control transport, and can give rise to different
states of confinement via transport bifurcation\citep{Kim:03a,Lothar:2012,burrell:97}.
Zonal flow shear often exhibits a finite autocorrelation time and
is not steady, as commonly assumed in simple shearing arguments. While
the mechanism of generation of zonal flows by turbulence is well established\citep{Chen:2000,diamond:05},
a central and unresolved question remains: \emph{what limits the coherence
and persistence of zonal shear in the saturated state} in the relevant
limit of weak collisional friction? Answering this question is essential
for a predictive understanding of turbulence regulation and transport
in magnetized plasmas.

In the magnetic fusion energy (MFE) context, the effectiveness of
a shearing process is determined both by its strength $\omega_{E}$
and by its persistence time i.e., autocorrelation time $\tau_{E}$.
Thus the shearing Kubo number $K_{u}\sim\omega_{E}\tau_{E}$ emerges
as a figure of merit of great interest. For $K_{u}\ge1$, the shearing
field is coherent and so the shearing damping decrement scales with
$\omega_{E}$ as most frequently assumed. For $K_{u}<1$, the shearing
field is stochastic, the shearing process is a random walk (in k-space)
and the shearing damping decrement scales with $\omega_{E}K_{u}$\citep{Kim:04}.
Thus, the physics which determines shear auto-coherence time $\tau_{E}$
is of great interest and relevance. Indeed, recent work\citep{Singh_2025}
has noticed that marked differences in zonal flows auto-coherence
occur along with differences in the state of turbulence and transport.
In particular, for otherwise similar conditions, zonal flows in ITG
turbulence for positive triangularity (PT) plasmas have much shorter
auto-coherence time $\tau_{E}$ than their counterparts in negative
triangularity (NT) plasmas. Concomitantly, transport and turbulence
levels in PT were shown to be larger than in NT. These observations
suggest that zonal shear coherence---rather than shear amplitude
alone---plays a decisive role in transport regulation.

\begin{figure}[H]
\includegraphics[scale=0.13]{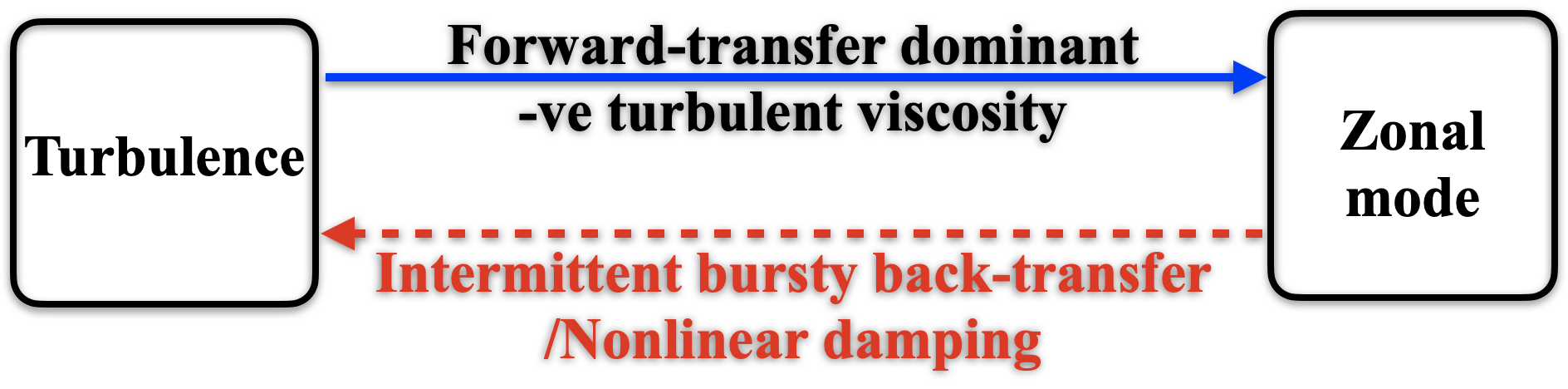}

\caption{Schematic of free energy transfer between turbulence and zonal mode.\label{fig:Schema1}}

\end{figure}

In this paper, we show that the zonal flow shear auto-coherence time
is limited by nonlinear spectral back transfer of free energy from
zonal mode to drift wave turbulence. This transfer combines with the
well known modulational process-- which couples excitation into the
zonal flow. Together, these two define the dynamic, saturated state
of zonal flows coexisting with drift wave turbulence in a collisionless
plasma. This is schematized in figure(\ref{fig:Schema1}). Nonlinear
back-transfer events are bursty, so zonal flows undergo an intermittent
birth and death process. We demonstrate that weaker (stronger) nonlinear
back transfer from zonal flows makes for lower (higher) levels of
transport and turbulence for fixed temperature gradient drive. 

Our analysis uses local gyrokinetic simulations of collisionless ITG
turbulence with adiabatic electrons performed using the GENE code\citep{Jenko_2000}.
We employ a subspace entropy-transfer diagnostic\citep{Nakata_2012,Maeyama_2021}
that explicitly measures nonlinear free energy exchange between turbulent
and zonal components. By examining a broad range of $a/L_{T}$, encompassing
both near-threshold and strongly driven regimes, we show that zonal
shear layers are highly dynamic objects: they are generated by transfer
from turbulence, evolve and merge, and ultimately decay through intermittent,
bursty events which return free energy from flows to turbulence. These
back-transfer events act as an effective nonlinear damping mechanism
that limits zonal shear coherence in the saturated state. Near marginal
stability, back-transfer is very weak, leading to long-lived, quasi-stationary
shear layers even as the absolute shear amplitude decreases, reflecting
the increasing relative importance of shear regulation close to threshold.
Results indicate that the \emph{zonal free energy back-transfer events
not only limit the coherence of zonal flow shear but also that of
all zonal corrugations. }Corrugations refer to the nonlinearly self-generated
zonal modes in density, temperature and other moments of the distribution
function. Shear layers and corrugations together constitute micro-barriers
in the plasma, forming radially localized structures that impede turbulent
transport through shear suppression and profile modulation\citep{GDP:2015}. 

Plasma shaping, in particular triangularity, allows a useful probe
on this fundamental mechanism. Several experiments\citep{Austin_2019,Marinoni_PoP_2019,Fontana_2017,Camenen_2007,Coda_2021,Marinoni_2021}
have reported improved confinement in NT over PT. A recent gyrokinetic
study\citep{Singh_2025} has shown that ITG transport reduction for
NT is due to turbulence auto-correlation length reduction\emph{ }due
to stronger nonlinearly generated zonal E\texttimes B shear as compared
to PT. By systematically comparing NT and PT equilibria, we show that
enhanced zonal shear in NT originates from reduced nonlinear back-transfer
from zonal modes to turbulence. \emph{Reduced back-transfer increases
the zonal shear autocorrelation time, thereby elevating the zonal
Kubo number and leading to more coherent, resilient, and long-lived
shear layers.} This outcome is independent of the absolute zonal kinetic
energy. In this sense, the NT--PT contrast is used as a controlled
setting that exposes the underlying physics: zonal shear coherency
is governed by nonlinear back-transfer, not by linear shear instability
criteria. As a result the effective shearing is stronger and hence
the turbulent heat diffusivity is lower for NT than for PT. Thus\emph{,
NT mitigates ITG turbulence and transport by extending the lifetime
and strength of the zonal ($E\times B$) shear through suppression
of these back-transfer events}. 

We note in passing that previous work\citep{rogers:00} has linked
the onset of collisionless zonal mode damping to essentially linear
instabilities of zonal modes, called tertiary modes. These occur at,
and define, the boundary of the so called Dimits shift (DS) regime\citep{Dimits_2000},
close to ITG marginal stability. There a linear stability approach
is applicable, due to the vanishingly low levels of turbulence in
the DS regime. While such an approach may identify the onset point
of zonal flow decay, it is intrinsically inapplicable to the broader
and more relevant problem of characterizing zonal flow properties
in a dynamic state of drift wave - zonal flow turbulence. In contrast,
the back-transfer--limited zonal shear coherency picture presented
here provides a fundamentally nonlinear framework for turbulence and
transport regulation, with potential implications for other zonal-flow--bearing
turbulence regimes in both laboratory and nature. \\
The remainder of this paper is organized as follows. Section(\ref{sec:stp})
presents the spatiotemporal evolution of zonal E\texttimes B shear
across temperature gradient and triangularity, establishing the qualitative
picture of enhanced coherence in NT relative to PT. Section(\ref{sec:lt})
quantifies the lifetime and radial size of zonal shear layers via
autocorrelation analysis. Section(\ref{sec:hd}) reports the variation
of turbulent heat diffusivity with temperature gradient and triangularity.
Section(\ref{sec:fzkes}) examines the zonal kinetic energy fraction
and effective shearing rate, revealing the counterintuitive low-ZKE
but high-shear character of NT. Section(\ref{sec:zkb}) defines and
evaluates the zonal Kubo number as a figure of merit for zonal shear
coherence. Section(\ref{sec:zet}) presents the central analysis of
this work: the zonal free energy transfer function $\mathcal{T}_{\text{zonal}}$,
including its mean, back-transfer statistics, and probability distributions
for NT and PT. Section(\ref{sec:pd}) connects zonal shear lifetime
to phase decorrelation via mean-squared displacement analysis of the
zonal shear phase, and suggests phase diffusion as a plausible microscopic
route by which back-transfer limits coherence. Section(\ref{sec:sum})
summarizes the main findings and their implications for reduced modeling
of drift-wave--zonal-flow turbulence. Supporting material, including
the spatiotemporal coherence of temperature and parallel velocity
corrugations, the derivation of the gyrokinetic free energy balance
equations, and the simulation parameters, is provided in the Appendices.

\section{Zonal $E\times B$ shear spatiotemporal pattern\label{sec:stp}}

\begin{figure}[H]
\includegraphics[scale=0.23]{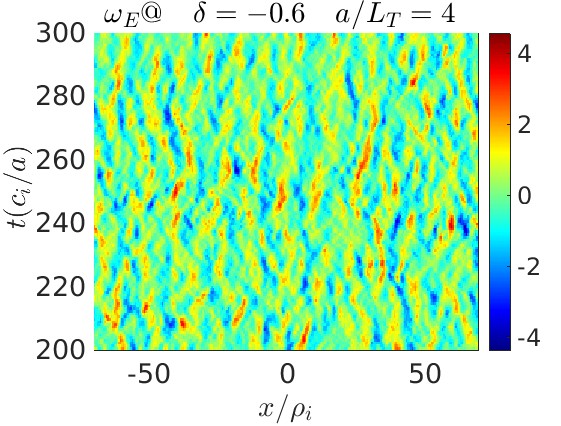}\includegraphics[scale=0.23]{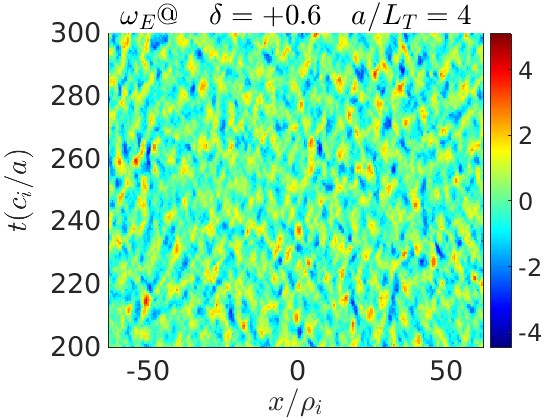}

\includegraphics[scale=0.23]{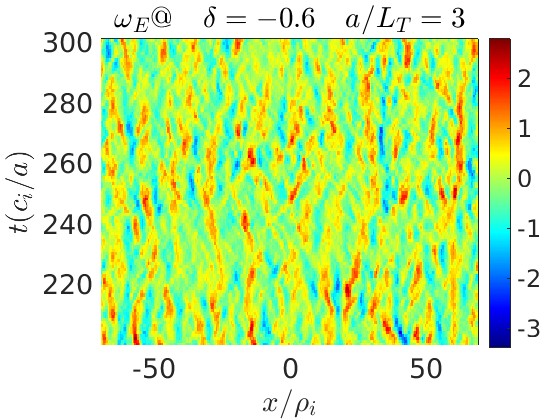}\includegraphics[scale=0.23]{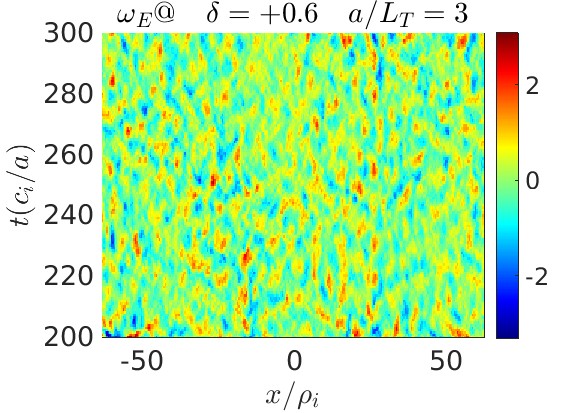}

\includegraphics[scale=0.23]{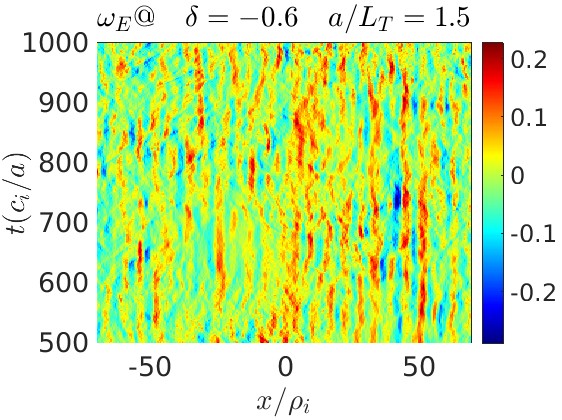}\includegraphics[scale=0.23]{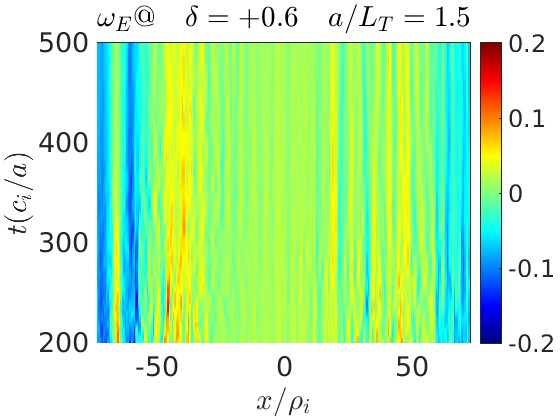}

\caption{Spatiotemporal pattern of zonal flow shear across triangularity and
temperature gradient. Notice the improvement in coherence as $a/L_{T}$
decreases and $PT\to NT$. \label{fig:stp}}
\end{figure}

The spatiotemporal evolution of zonal flow shear across temperature
gradient $a/L_{T}$ for both PT and NT cases with triangularity $\delta=\pm0.6$
is shown in figure(\ref{fig:stp}). The shear layers are highly dynamic,
exhibiting continuous cycles of formation, maturation, radial migration,
and eventual decay. These plots demonstrate the following key features.
The shearing pattern is \emph{visibly} more coherent in NT than in
PT. The shearing pattern gradually becomes more coherent with decreasing
$a/L_{T}$. Near marginality, the coherence increases drastically--the
shear layers become remarkably stationary and long-lived. As expected,
the absolute strength of zonal shear $\omega_{E}$ decreases with
decreasing $a/L_{T}$, consistent with the reduced turbulence drive.
Thus, these spatiotemporal plots highlight that NT favors more coherent
and long-lived shear structures than PT, particularly near threshold.
The coherence of shear layers is quantified by calculating the spatiotemporal
autocorrelation functions. The spatiotemporal pattern of corrugations
also become more coherent approaching marginality, as shown in figure(\ref{fig:stp-1})
in the Appendix(\ref{sec:corrg}). 

\section{Life time and size of zonal shear\label{sec:lt}}

\begin{figure}[H]
\includegraphics[scale=0.23]{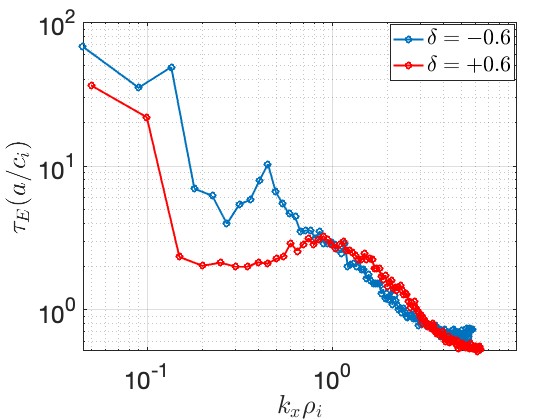}\includegraphics[scale=0.23]{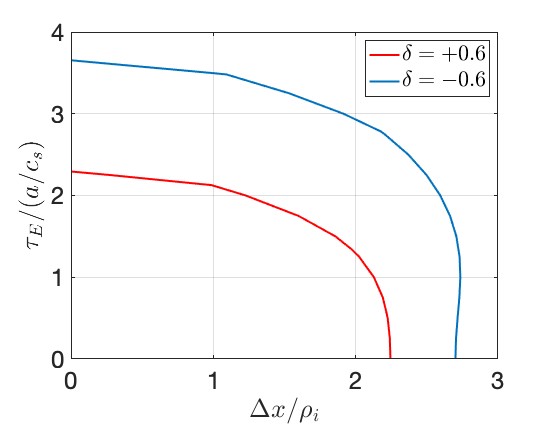}

\includegraphics[scale=0.23]{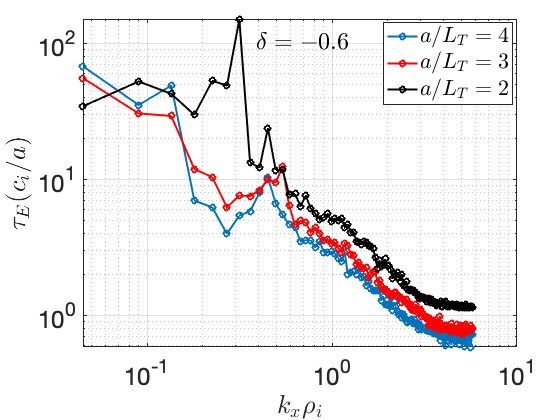}\includegraphics[scale=0.23]{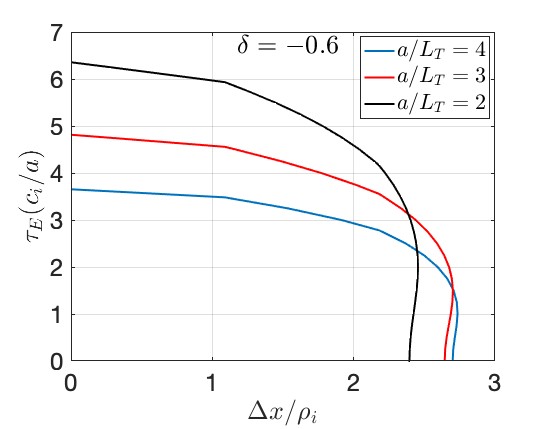}

\caption{Top left: The $k_{x}$- spectra of the life time ($1/e$ time of the
auto-correlation function) of zonal shear for PT and NT. Top right:
life time and radial size of the zonal shear obtained from $1/e$
contour level of the 2d autocorrelation function in $(x-t)$ space.
Bottom left: The $k_{x}$- spectra of the life time ($1/e$ time of
the auto-correlation function) of zonal shear for NT for varying $a/L_{T}$.
Bottom right: life time and radial size of the zonal shear obtained
from $1/e$ contour level of the 2d autocorrelation function in $(x-t)$
for various $a/L_{T}$. \label{fig:zsac}}
\end{figure}

The lifetime of zonal shear is determined from the exponential decay
time of the temporal autocorrelation function, while its radial scale
is obtained from the exponential decay width of the radial autocorrelation
function. Figure(\ref{fig:zsac}) shows the radial wavenumber spectra
of zonal shear lifetime, together with the joint dependence of lifetime
and radial size extracted from the $1/e$-contour level of the two-dimensional
autocorrelation in $(x,t)$ for $\delta=\pm0.6$. These results reveal
that the\textbf{ }zonal shear lifetime is systematically longer for
NT than for PT, indicating more persistent shear structures in NT
plasmas.\textbf{ }The radial extent\textbf{ }of zonal shear is also
larger for NT than for PT, implying broader shear layers.\textbf{
}NT, therefore, supports zonal shear layers that are both broader
and more temporally persistent, leading to enhanced spatiotemporal
coherence as compared to PT. Similarly, the autocorrelation analysis
across $a/L_{T}$ reveals the emergence of the following trends.\textbf{
}Zonal shear lifetime increases at all radial scales as $a/L_{T}$
decreases, i.e., as the system approaches marginal stability. The
radial extent of shear structures also increases with decreasing $a/L_{T}$.
These quantitative results are consistent with the previous observations,
where the spatiotemporal pattern is shown to become increasingly coherent
near marginality. Together, they demonstrate that \emph{lower drive
conditions favor more coherent, longer-lived, and spatially extended
zonal shear layers}. 

\section{Turbulent heat diffusivities\label{sec:hd}}

\begin{figure}[h]
\includegraphics[scale=0.47]{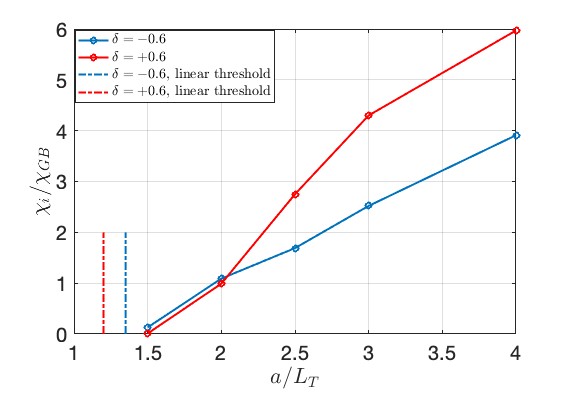}

\caption{Left: Turbulent heat diffusivity vs $a/L_{T}$. The vertical lines
mark the linear thresholds. \label{fig:hd} }
\end{figure}

Figure(\ref{fig:hd}) shows the variations of turbulent heat diffusivity
with $a/L_{T}$ for $\delta=\pm0.6$. For $a/L_{T}>2$, the heat diffusivity
is systematically lower in NT than in PT, indicating improved confinement
in NT shaping away from marginality. As marginality is approached,
the difference between NT and PT narrows. Notice that the Dimits shift
is less pronounced in NT than in PT. This reduction is due to the
fact that the nonlinear critical gradients are comparable in PT and
NT, while the linear critical gradient is higher in NT. Consequently,
NT provides a confinement advantage away from marginality, but not
near threshold. The Dimits shift is widely understood to originate
from the suppression of turbulence by self-generated zonal flows;
its magnitude therefore encodes how effectively zonal flows can suppress
transport in different shaping configurations. The reduction of linear
growth rates and the increase of linear critical gradients in NT are
linked to the reduced eigenmode-averaged magnetic drift frequency
in NT shaping compared to PT. To extract the microphysics of these
macroscopic trends, we next analyze the variations of fluctuation
kinetic energy, zonal kinetic energy, and zonal $E\times B$ shear
across $a/L_{T}$ and $\delta$ in the following. 

\section{Zonal kinetic energy and zonal $E\times B$ shear\label{sec:fzkes}}

\begin{figure}[H]
\includegraphics[scale=0.47]{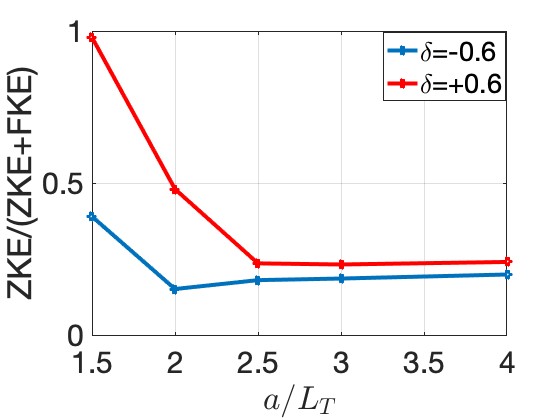}

\includegraphics[scale=0.47]{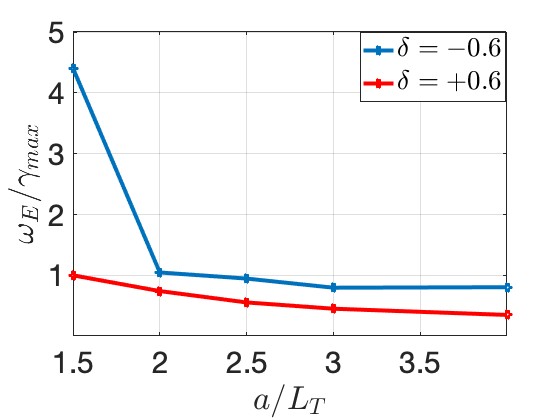}

\caption{Zonal kinetic fraction and effective shearing rates across temperature
gradient for PT and NT. \label{fig:zet-1-1}}
\end{figure}

Fig(\ref{fig:zet-1-1}) shows the dependence of zonal kinetic energy
(ZKE) fraction $ZKE/(ZKE+FKE)$, {[}mathematical expression for $ZKE$
and $FKE$ are given in the Appendix(\ref{sec:NM}){]} on $a/L_{T}$.
It also shows the effective shearing rate $(\omega_{E}/\gamma_{max})$
i.e., the ratio of $E\times B$ shearing rate to the maximum linear
growth rate vs the temperature gradient $a/L_{T}$ for both PT and
NT cases. The results reveal the following systematic trends. The
ZKE-fraction is consistently higher for PT than NT across all $a/L_{T}$.
Away from marginality, the ZKE-fraction varies only weakly with $a/L_{T}$
. However, it rises sharply when approaching marginality from above,
with PT exhibiting a much more pronounced increase than NT. The normalized
shearing rate $\omega_{E}/\gamma_{max}$ increases as marginality
is approached from above, reflecting the increasing relative importance
of shear in regulating turbulence near threshold. We emphasize that,
\emph{the effective shearing rate is consistently higher, while ZKE-fraction
is lower, for NT than for PT. }This trend persists across all $a/L_{T}$.
This finding not only clarifies the relationship between flow and
fluctuation energetics across both $a/L_{T}$ and $\delta$, but also
challenges the commonly held assumption that high zonal kinetic energy
necessarily corresponds to strong zonal shear. Instead, NT emerges
as a \emph{low-ZKE but high-shear state}, while PT emerges as a \emph{high-ZKE
but low-shear state}. Since the ability of zonal flows to regulate
turbulence depends directly on the shearing rate than on the absolute
zonal kinetic energy, these results emphasize that effective shear,
rather than zonal energy, is the critical quantity governing confinement
trends. 

\section{Zonal Kubo number\label{sec:zkb}}

\begin{figure}[H]
\includegraphics[scale=0.47]{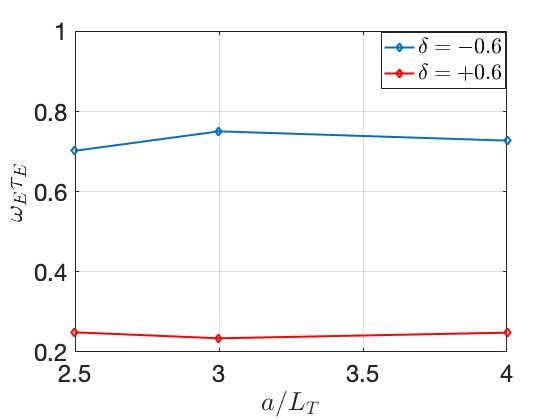}

\caption{Zonal Kubo number vs $a/L_{T}$. \label{fig:zkb}}
\end{figure}

The zonal Kubo number is defined as the product of the r.m.s zonal
ExB shearing rate and the zonal ExB shear autocorrelation time i.e.,
$K_{u}=\omega_{E}\tau_{E}$. It provides a figure of merit for the
coherence of the zonal flow. Thus $K_{u}>1$ implies coherent shearing
effects. $K_{u}<1$ indicates stochastic effective shearing effects.
Since the beneficial effects of negative triangularity (NT)---namely
reduced turbulent transport and improved confinement relative to positive
triangularity (PT)---emerge only well above marginal stability (
$a/L_{T}\ge2.5$), we evaluate $K_{u}$ in this strongly driven regime.
Figure(\ref{fig:zkb}) shows that $K_{u}<1$ for both PT and and NT,
suggesting stochastic shearing in both cases. However, $K_{u}$ is
systematically larger in NT than in PT and is substantial. Thus, zonal
shear is more temporally coherent in NT, rendering shear suppression
more effective that is consistent with the reduced turbulence and
transport observed there. The physical origin of the enhanced $K_{u}$
in NT is identified and characterized systematically in the following.
\textbf{ }

\section{Zonal free energy transfer analysis \label{sec:zet}}

\begin{figure}[h]
\includegraphics[scale=0.22]{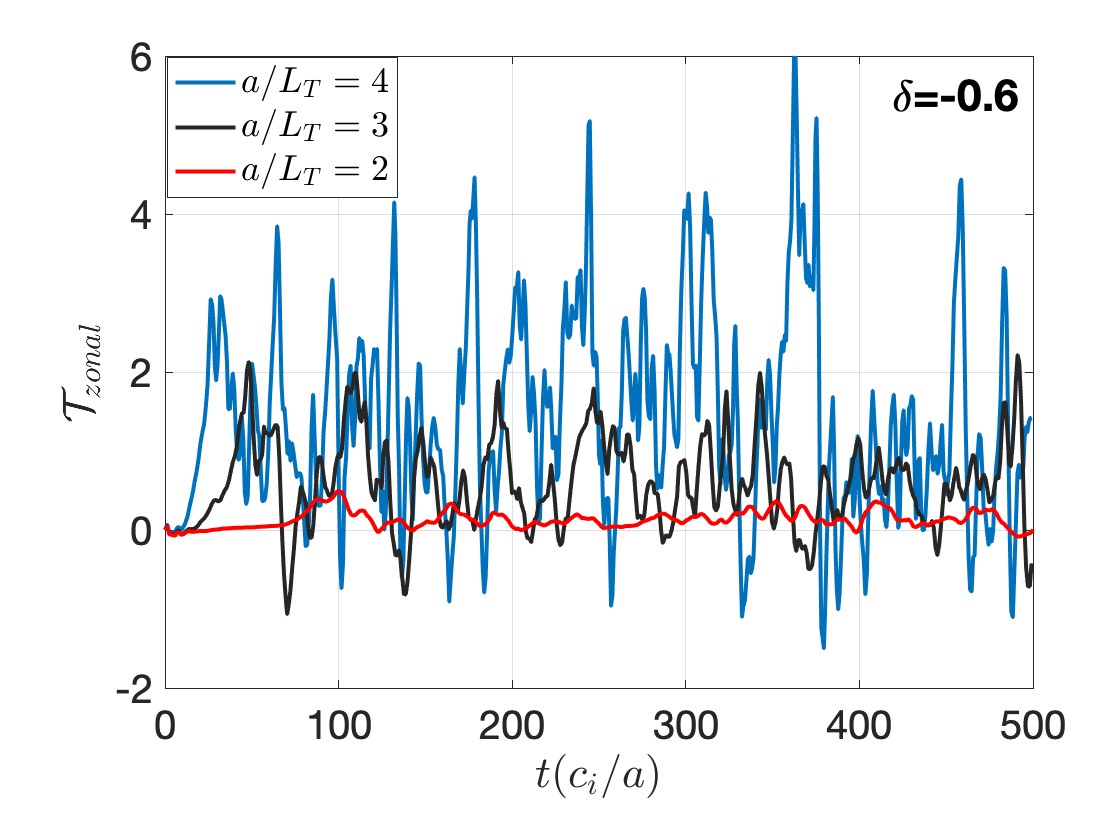}

\includegraphics[scale=0.18]{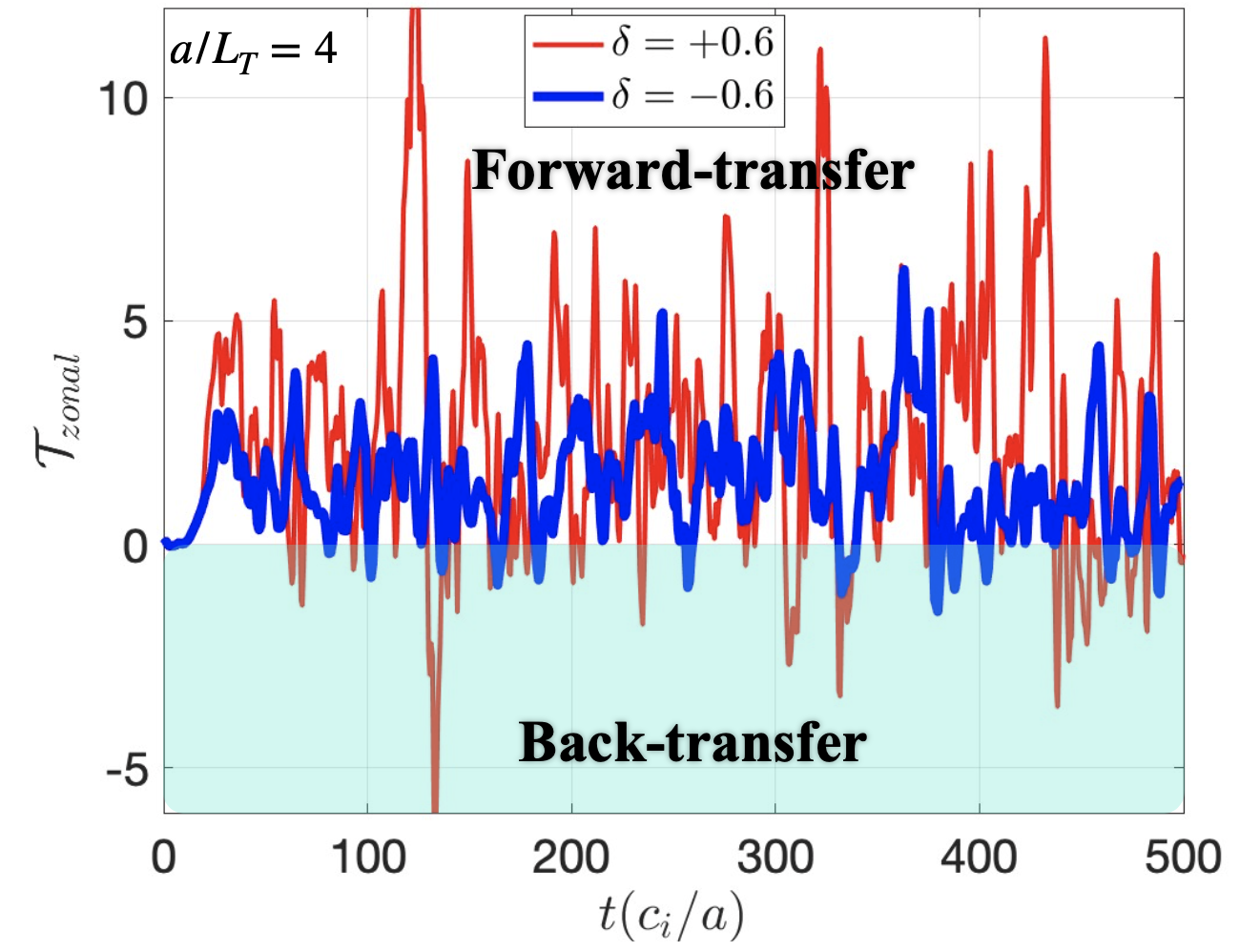}

\caption{Temporal evolution of zonal free energy transfer rates across triangularity
and temperature gradient.\label{fig:zet-2}}
\end{figure}

Previous sections discussed on the behavior of zonal shear coherence,
strength and energy on $a/L_{T}$ and $\delta$. This raises the question
of what physics accounts for these trends. To address this, we investigated
the zonal free energy transfer function $\mathcal{T}_{zonal}$, defined
explicitly in Equation(\ref{eq:etf}) in the Appendix(\ref{sec:NM}).
$\mathcal{T}_{zonal}$ is the zonal component of the triad transfer
function. $\mathcal{T}_{zonal}$ provides direct information about
the exchange of free energy between turbulence and zonal modes, and
thereby about the mechanisms that set the strength and coherence of
zonal shear and corrugations across different regimes. Notice that
the free energy is a nonlinear invariant, analogous to the phase space
enstrophy. Figure(\ref{fig:zet-2}) presents the temporal evolution
of zonal free energy transfer rates. Notice that, on average, there
is net transfer of free energy from turbulent fluctuations to zonal
modes. This shows that the turbulence drives and sustains zonal flows.
At the same time, intermittent excursions into negative values reveals
episodes of back-transfer, in which free energy flows from the zonal
modes back into the turbulence. We present evidence that these \emph{back-transfer
events are closely associated with the stability properties of the
zonal mode and effectively limit both the strength and the lifetime
of zonal flow shear} \emph{and corrugations}. 

As $a/L_{T}$ decreases, both the amplitude of the forward transfer
and the magnitude of back-transfer are reduced. Furthermore, a direct
comparison between the PT and NT cases shows that the transfer amplitude
and the extent of back-transfer are consistently smaller for NT than
for PT. Together, these observations point to significant differences
in the energy dynamics of zonal flows across regimes, with clear implications
for the persistence and regulating role of zonal shear. In the following,
we characterize these behaviors systematically by analyzing the mean
transfer and the statistics of back-transfer events.

\subsection{\emph{Mean-zonal-transfer}}

The mean zonal transfer rate, defined as the time-averaged zonal free
energy transfer, is consistently positive, indicating that on average
free energy flows from turbulence into the zonal mode. Several systematic
trends emerge: The time-averaged zonal transfer rate increases with
increasing $a/L_{T}$. This trend explains why both the zonal kinetic
energy and the absolute zonal shear increase with $a/L_{T}$. The
time-averaged zonal transfer rate is systematically lower for NT than
for PT. This is consistent with the observation that the zonal kinetic
energy is also lower for NT compared to PT. The $k_{x}$-spectra of
$\mathcal{T}_{zonal}$ show that the mean energy transfer from turbulence
to the zonal mode decreases across all radial scales as $a/L_{T}$
approaches marginal stability from above. Similarly, \emph{the spectra
indicate a systematic reduction in transfer across all radial scales
when transitioning from PT to NT}. Consequently, the total the zonal
kinetic energy(ZKE) is lower for NT than for PT. 

\begin{figure}[h]
\includegraphics[scale=0.23]{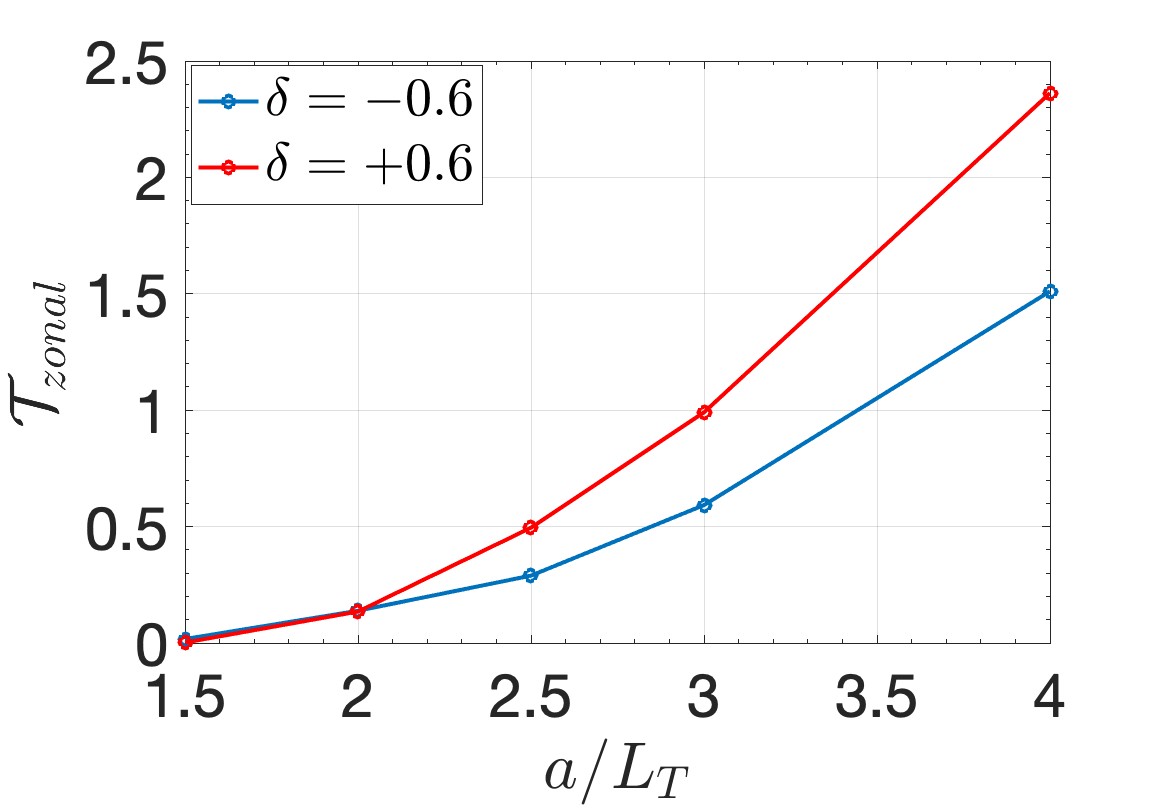}

\includegraphics[scale=0.47]{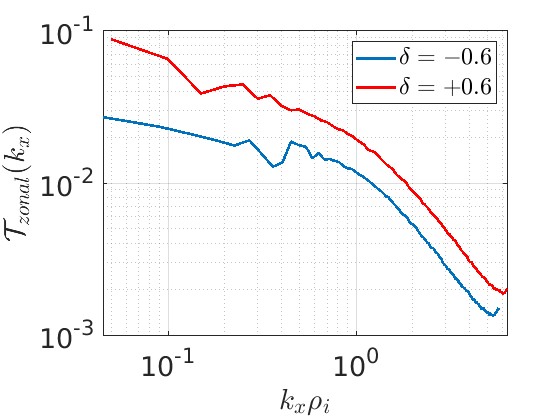}

\caption{Mean zonal free energy transfer rates spectra and mean zonal free
energy transfer across temperature gradient\label{fig:zet}}
\end{figure}

\subsection{\emph{Zonal back-transfer (limits zonal shear coherency)}}

\begin{figure}[H]
\includegraphics[scale=0.24]{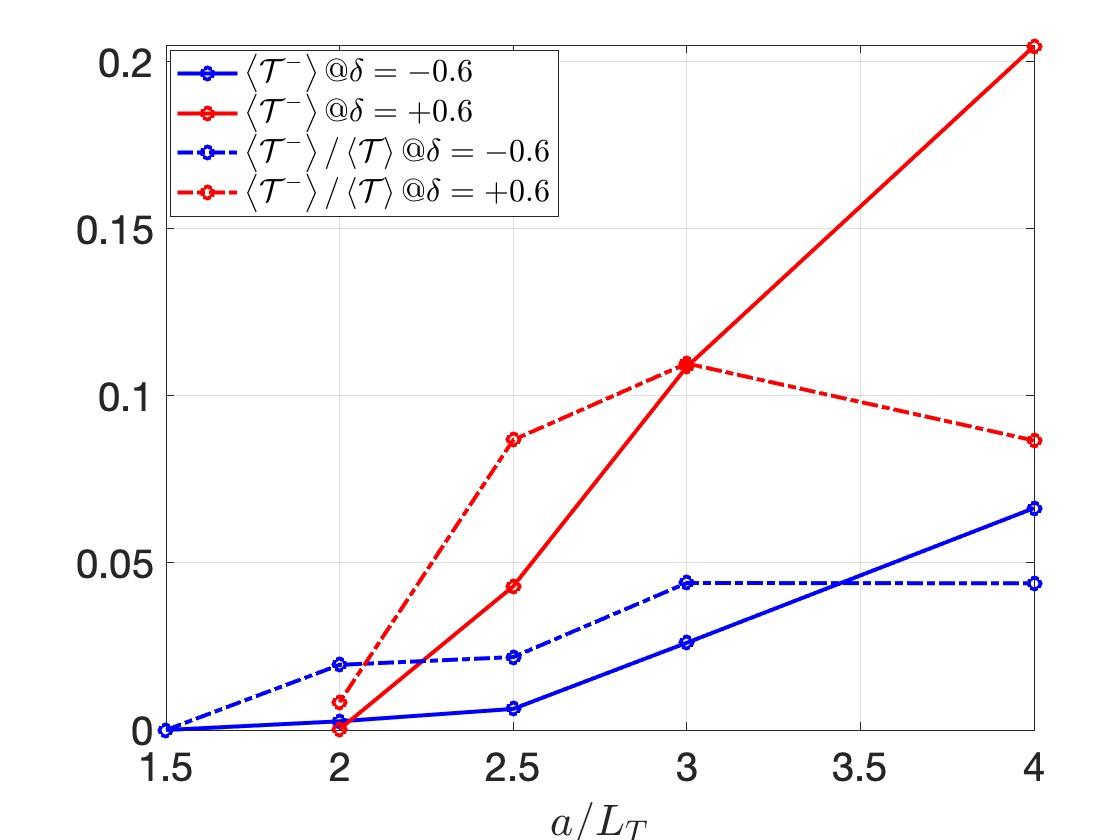}

\caption{Quantification of zonal-back-transfer across temperature gradient
for NT and PT shapes. \label{fig:zet-1}}
\end{figure}

\begin{figure}[H]
\includegraphics[scale=0.47]{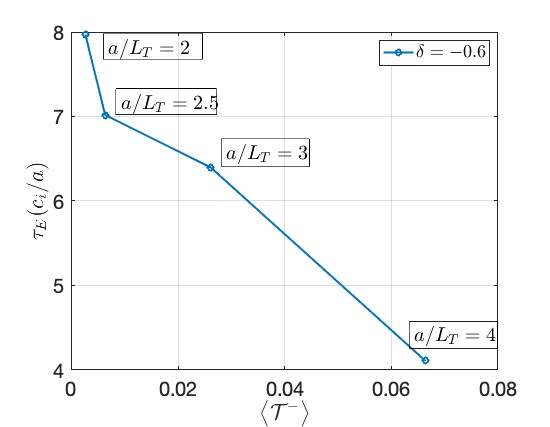}

\caption{Zonal shear life-time vs back transfer. \label{fig:zet-1-2}}
\end{figure}

Variations of back-transfer with temperature gradient are shown in
figure(\ref{fig:zet-1}). Here, the back-transfer of energy from zonal
mode to fluctuations is quantified in two complementary ways: a) the
\emph{mean-back-transfer}\textbf{,} defined as temporal average over
negative transfer events $\left\langle \mathcal{T}^{-}\right\rangle =\frac{\int dt\mathcal{T}^{-}}{\int dt}$,
and b) the\emph{ back-transfer-fraction,} defined as ratio of time-averaged
negative transfer over total time-averaged transfer $\frac{\left\langle \mathcal{T}^{-}\right\rangle }{\left\langle \mathcal{T}\right\rangle }=\frac{\int dt\mathcal{T}^{-}}{\int dt\mathcal{T}}$.
Clearly, the back-transfer increases with the temperature gradient
$a/L_{T}$. As a consequence, the zonal shear lifetime is shorter
at higher temperature gradients. Near marginal stability, however,
the back-transfer fraction is significantly reduced, leading to a
pronounced enhancement of zonal flow coherence. A similar trend has
been reported previously regarding back-transfer of Reynolds power
with external drive in a forced Hasegawa-Mima system\citep{RD:2024}.
The back-transfer fraction is lower for NT than for PT, for fixed
$a/L_{T}$. This implies a more resilient and coherent shear for NT
than for PT. The direct relation between zonal shear life time and
zonal back-transfer is quantified in figure(\ref{fig:zet-1-2}). Note
that back-transfer events also limit the coherency of corrugations.
The corrugations life-time is graphed against the back-transfer in
figure(\ref{fig:cltbt}) and confirms this claim. This is discussed
further in figures(\ref{fig:stp-1}) in the Appendix(\ref{sec:corrg}).
We show that the improved coherence of corrugations is due to suppression
of zonal back-transfer events near marginality. Thus, these results
identify \emph{back-transfer as the nonlinear mechanism limiting zonal
flow (and corrugations) persistence} \emph{and coherence}. Its reduction
near marginality and in strongly NT configurations enhances zonal
shear coherence and lifetime, thereby strengthening turbulence regulation.
This underscores the fundamental role of free energy transfer dynamics
in setting the balance between turbulence drive and zonal flow saturation.
\begin{figure}[H]
\includegraphics[scale=0.47]{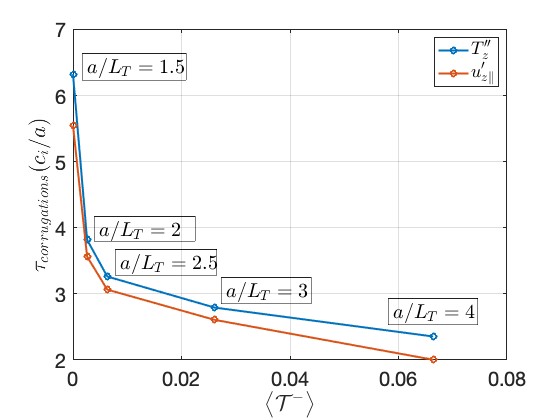}

\caption{Life-time of corrugations vs zonal back-transfer. \label{fig:cltbt}}
\end{figure}

\subsection{\emph{Statistics of zonal free energy transfer}}

\begin{figure}[H]
\includegraphics[scale=0.47]{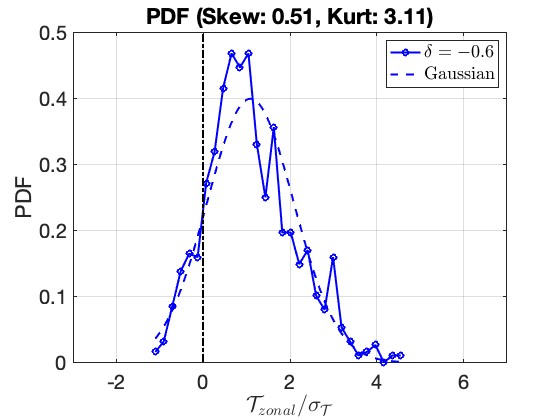}

\includegraphics[scale=0.47]{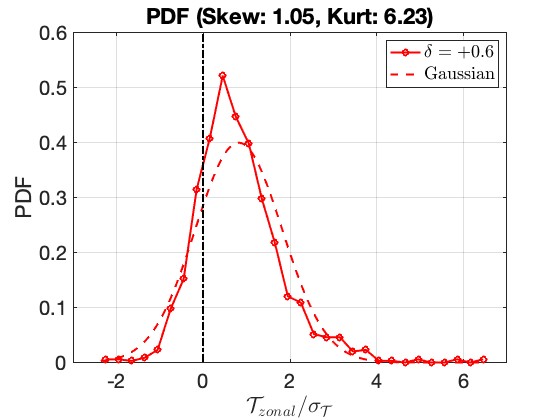}

\caption{PDF of zonal free energy transfer amplitudes.\label{fig:PDF}}
\end{figure}

All statistical analyses are performed on the time series of $\mathcal{T}_{zonal}$
(t) in the fully saturated state, defined by $tc_{i}/a\ge100$. This
threshold is chosen to exclude the initial linear growth and early-nonlinear
transient phases, ensuring that the PDFs, waiting-time distributions,
and derived intermittency statistics characterize the statistically
stationary drift-wave--zonal-flow system rather than its approach
to saturation. The probability distribution function (PDF) of the
instantaneous zonal free energy transfer amplitudes for NT and PT
are shown in figure (\ref{fig:PDF}). Both PDFs exhibit non-Gaussian
features with Kurtosis and Skewness deviating from the Gaussian baseline,
consistent with bursty, intermittent dynamics. The ratio of the standard
deviation $\sigma_{\mathcal{T}}$ to the mean of $\mathcal{T}_{\text{zonal}}$
i.e., $\sigma_{\mathcal{T}}/\left\langle \mathcal{T}_{zonal}\right\rangle $
also provides a simple quantitative measure of intermittency. For
NT: $\sigma_{\mathcal{T}}/\left\langle \mathcal{T}_{zonal}\right\rangle =0.92$,
Skewness$=0.51$, Kurtosis$=3.11$ and the fraction of time spent
in the back-transfer state$=12.33\%$. The corresponding PT values
are markedly larger: $\sigma_{\mathcal{T}}/\left\langle \mathcal{T}_{zonal}\right\rangle =1.22,$
Skewness$=1.05$, Kurtosis$=6.23$ and the back-transfer residence
time $=18.28\%$. Taken together these statistics demonstrate that
zonal free-energy transfer in NT is significantly less intermittent
and closer to Gaussian, with a back-transfer residence time approximately
$33\%$ shorter than in PT. 
\begin{figure}[H]
\includegraphics[scale=0.47]{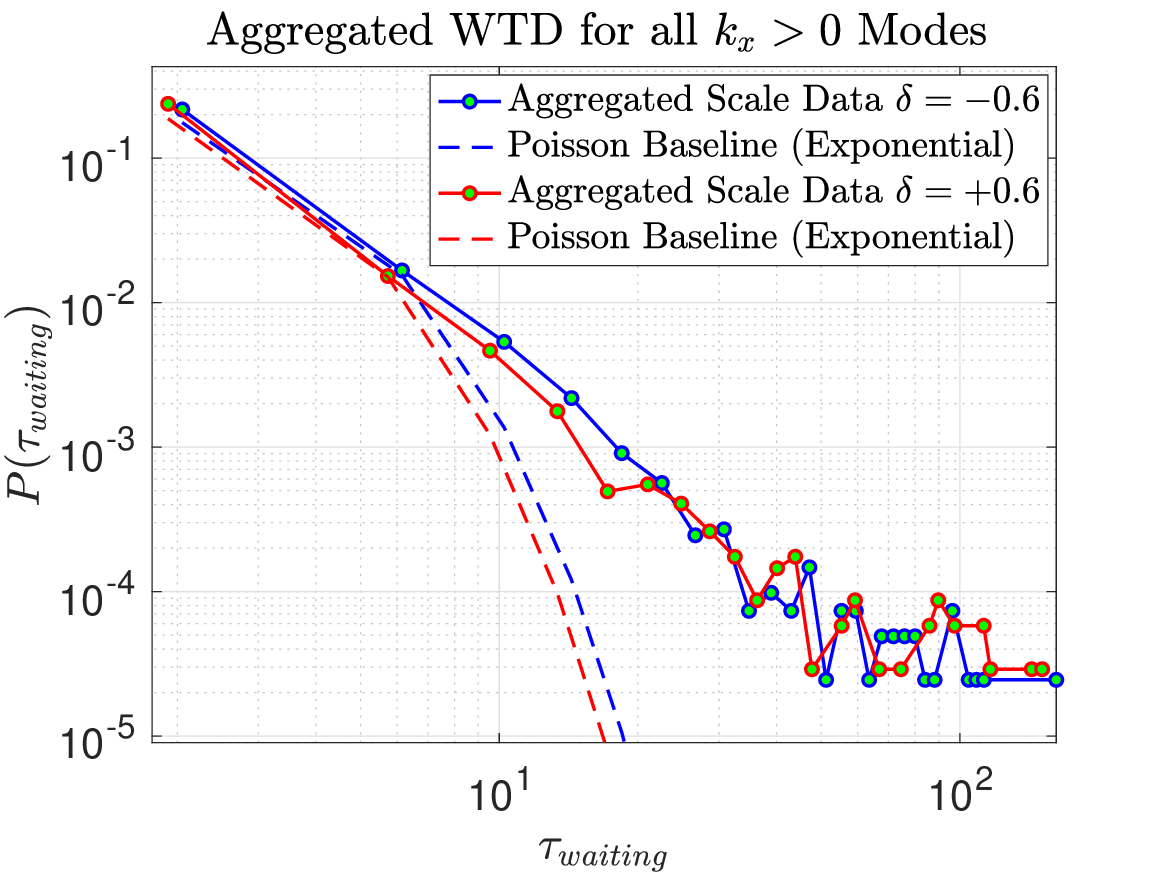}

\caption{Aggregated PDF of the waiting times of zonal back transfer onset.\label{fig:WTD}}
\end{figure}
To characterize the temporal organization of back-transfer events,
we compute the waiting-time distribution (WTD) of their onset for
both configurations. Back-transfer events are identified from the
$\mathcal{T}_{zonal}$ time series by the criterion $\mathcal{T}_{zonal}<0$
, and the inter-event waiting times are aggregated over all $k_{x}>0$
modes. The aggregated WTDs are constructed from $N=9948$ (NT) and
$N=9021$ ( PT) events respectively, providing robust statistics across
the bulk of each distribution; the far tail is populated by fewer
events and is subject to larger sampling uncertainty. The resulting
WTDs are shown in Figure(\ref{fig:WTD}). Both distributions exhibit
heavy tails, indicating that back-transfer bursts are not Poisson-distributed
in time but instead arrive in clusters --- a hallmark of intermittent,
self-organized dynamics. This clustering is quantified by the burstiness
parameter\citep{Goh_2008} $B=\frac{C_{V}-1}{C_{V}+1}$, where $C_{V}=\frac{\sigma_{\tau_{w}}}{\left\langle \tau_{w}\right\rangle }$
is the coefficient of variation of the WTD. $\sigma_{\tau_{w}}$ is
the standard deviation of waiting times and $\left\langle \tau_{w}\right\rangle $
is the mean waiting time. We obtain $B=+0.52\pm0.02$ for NT and $B=+0.56\pm0.02$
for PT with bootstrap uncertainties\citep{Efron1979} (n=1000 resamples),
both far exceeding the Poisson value $B=0$, confirming strongly clustered,
non-Poissonian back-transfer in both configurations. The difference
between NT and PT is marginal, with $B$ for PT higher than for NT,
consistent with other strong indicators (kurtosis, skewness, normalized
standard deviation, residence time) of enhanced intermittency in PT.
The NT distribution is shifted toward longer waiting times relative
to PT, quantifying the reduced frequency of back-transfer onset in
the more coherent NT configuration. To assess whether consecutive
inter-event times carry memory of one another, we compute the memory
coefficient $M$ --- the Pearson correlation between successive waiting
times\citep{Goh_2008}. We find $M=+0.2\pm0.01$ for NT and $M=+0.19\pm0.01$
for PT (bootstrap $1\sigma$), both $19\sigma$ above zero, establishing
that the back-transfer event sequence is not a pure renewal process:
a long quiescent interval tends to be followed by another long one,
reflecting the slow modulation of back-transfer susceptibility by
the evolving zonal mode amplitude. Crucially, $M$ is statistically
indistinguishable between NT and PT, demonstrating that the NT--PT
contrast resides entirely in the marginal WTD --- the frequency and
intensity of burst arrivals --- rather than in the sequential correlation
structure of waiting times.

Collectively, these results establish that NT exhibits both lower
back-transfer burst amplitudes and longer inter-burst waiting times
compared to PT. The combination of reduced burst amplitude (lower
$\sigma_{\mathcal{T}}$ , skewness, and kurtosis) and sparser burst
arrival (longer waiting times, lower back-transfer residence time),
with statistically similar back-transfer burstiness $B$ and inter-event
times memory $M$ provides a quantitative, statistically grounded
account of why NT sustains longer zonal shear coherence times.

\section{Connection of zonal shear life time with Phase decorrelation\label{sec:pd}}

The mechanism by which back-transfer shortens $\tau_{E}$ operates
through stochastic phase scattering: turbulent eddies that receive
free energy from the zonal mode re-emit wave packets that interfere
destructively with the coherent zonal phase, eroding its temporal
coherence. This is the phase-diffusion picture of zonal flow decorrelation. 

This picture is quantified through the mean-squared displacement (MSD)
of the zonal shear phase $\varphi_{k_{x}}$, $\text{MSD}(\tau)=\left\langle \left[\varphi_{k_{x}}(t+\tau)-\varphi_{k_{x}}(t)\right]^{2}\right\rangle _{t}$,
which is systematically smaller for NT than for PT at every radial
scale $k_{x}$. A smaller MSD implies slower phase decoherence, a
reduced phase diffusion coefficient $D_{\varphi}=\frac{1}{2}\frac{d\text{MSD}}{dt}$,
and consequently a longer phase diffusion time $\tau_{\varphi}=\frac{1}{D_{\varphi}}$. 

\begin{figure}[H]
\includegraphics[scale=0.47]{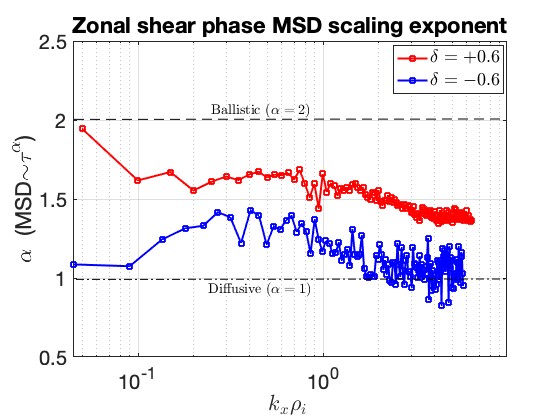}

\caption{Scaling exponents of MSD of zonal shear phase.\label{fig:msd}}
\end{figure}

The nature of the phase dynamics is further characterized by the MSD
scaling exponent $\alpha$, defined by $\text{MSD}\sim\tau^{\alpha}$:
$\alpha>1$ indicates superdiffusion, $\alpha=1$ pure diffusion,
and $\alpha<1$ subdiffusion. Figure(\ref{fig:msd}) compares $\alpha(k_{x})$
for NT and PT. In PT, the phase dynamics is superdiffusive across
all radial scales, consistent with the stronger back-transfer activity
in PT. In NT the situation is qualitatively different: dominant low-$k_{x}$
modes remain superdiffusive, but subdominant large-$k_{x}$ modes
approach pure diffusion, suggesting weaker phase scattering at fine
radial scales, in line with the reduced back-transfer in NT. 

The phase MSD is generally not linear in the time lag. At large lags
it also shows oscillation and saturation, likely due to finite sample
effect. The local quantity $D_{\varphi}$ is therefore lag-dependent
in general. We obtain an effective, lag-independent coefficient by
a linear least-squares fit of the MSD over the early-lag range $\tau c_{i}/a\in[0,33]$,
avoiding the late oscillatory region. Note that this time window is
bigger that the global autocorrelation time of zonal shear for most
$k_{x}$ and thus safely captures the phase decoherence dynamics.
The quantity $1/D_{\varphi}$ should thus be read as an effective
coherence time, set by the initial rate of phase spreading, rather
than as a diffusion time in the strict sense. We acknowledge that
$1/D_{\varphi}$ may over- or underestimate the true coherence time
at a given $k_{x}$. 
\begin{figure}[H]
\includegraphics[scale=0.47]{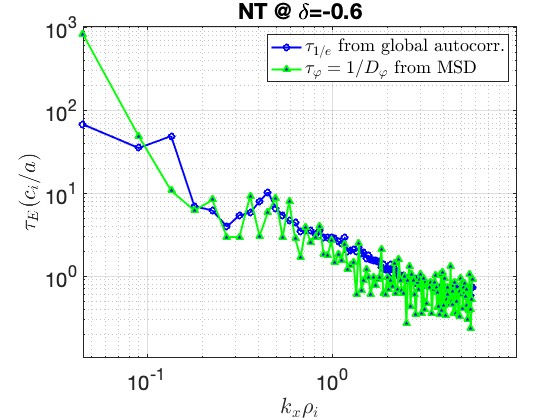}

\includegraphics[scale=0.47]{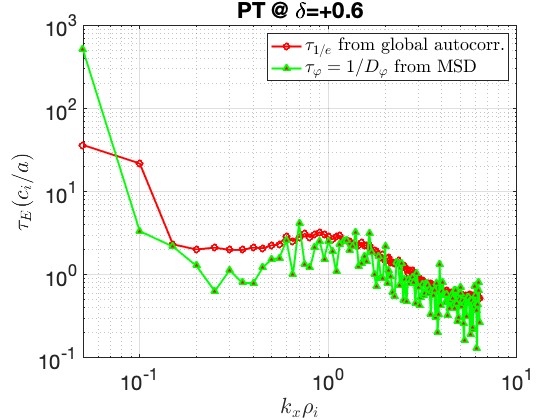}

\caption{Zonal shear life time from autocorrelation function and phase diffusion.\label{fig:tauE}}
\end{figure}
Figure(\ref{fig:tauE}) compares the zonal shear lifetime estimated
from the global autocorrelation function ($\tau_{1/e}$ ) with the
phase diffusion time $\tau_{\varphi}=\frac{1}{D_{\varphi}}$. The
qualitative agreement of the $k_{x}$-trend between these independent
estimates for both configurations indicates that zonal shear coherence
is closely tied to phase decorrelation process. Residual discrepancies
between the two estimates arise from three sources: non-Gaussian statistics
of the phase increments (excess kurtosis of the $\Delta\varphi$ distribution),
the finite decorrelation time of the shear amplitude envelope, and
the covariance between amplitude and phase. Crucially, the mismatch
is larger for PT than for NT, implying that these non-Gaussian and
amplitude-phase coupling effects are more prominent in PT. This is
consistant with the stronger, more intermittent back-transfer bursts
in PT, which would be expected to disrupt both the amplitude and the
phase coherence of the zonal shear --- an effect that the scalar
$D_{\varphi}$ alone cannot fully capture. These results and the statistical
properties of the transfer dynamics are further summarized in the
following table. Taken together, the analyses in Sections(\ref{sec:zet})
and (\ref{sec:pd}) lead us to speculate that back-transfer events
enhance the phase scattering of the zonal shear, providing a possible
microscopic, mode-resolved route to a finite $\tau_{E}$. A direct
event-conditioned test of this picture is left for future work.
\begin{table}[h]
\begin{tabular}{|>{\raggedright}p{2cm}|>{\centering}p{2cm}|>{\centering}p{2cm}|>{\centering}p{2cm}|}
\hline 
Metric & PT & NT & Implications\tabularnewline
\hline 
\hline 
Normalized std. dev. of transfer $\sigma_{\mathcal{T}}/\left\langle \mathcal{T}_{zonal}\right\rangle $ & 1.22 & 0.92 & NT transfer is less volatile\tabularnewline
\hline 
PDF skewness of $\mathcal{T}_{zonal}$ & 1.05 & 0.51 & NT closer to symmetric / Gaussian\tabularnewline
\hline 
PDF  kurtosis of $\mathcal{T}_{zonal}$ & 6.23 & 3.11 & NT less intermittent\tabularnewline
\hline 
Back-transfer residence time & 18.28\% & 12.33\% & NT spends $\sim33\%$ less time in back-transfer state\tabularnewline
\hline 
Back-transfer Waiting-time distribution & Shorter inter-burst gaps  & Longer inter-burst gaps & NT back-transfer less frequent\tabularnewline
\hline 
MSD scaling exponent $\alpha(k_{x})$ & Superdiffusive at all $k_{x}$ & Superdiffusive at low $k_{x}$; approaches diffusion at large $k_{x}$ & Weaker phase scattering at fine radial scales in NT\tabularnewline
\hline 
Phase diffusion coefficient $D_{\varphi}$ & Higher & Lower & Slower phase decoherence in NT\tabularnewline
\hline 
Coherence time agreement $\tau_{1/e}$ vs $1/D_{\varphi}$ & Larger mismatch & Closer agreement & Stronger amplitude--phase coupling and non-Gaussian effects in PT
due to intense bursts\tabularnewline
\hline 
\end{tabular}

\caption{Summary of statistical properties of the free energy transfer dynamics
and the zonal shear phase dynamics for NT and PT.}
\end{table}

\section{Summary\label{sec:sum}}

Back-transfer of zonal free energy is demonstrated to be a major nonlinear
mechanism limiting zonal mode coherency in collisionless gyrokinetic
ITG turbulence with adiabatic electrons. While the net free energy
transfer is from turbulence to zonal modes, intermittent bursty back-transfer
events return energy from the zonal flows to the turbulence, thereby
limiting the stability and spatiotemporal coherence of zonal shear
in the saturated state. This suggests that zonal flow saturation is
governed by an effective nonlinear damping process linked to back-transfer
events. The microscopic mechanism by which back-transfer limits the
zonal shear autocorrelation time $\tau_{E}$ operates predominantly
through stochastic phase scattering. The effective phase diffusion
time $\tau_{\varphi}$ inferred from the MSD of the zonal shear phase
exhibits a trend qualitatively similar to that of $\tau_{E}$ across
radial scales, showing that the zonal shear lifetime is closely tied
to phase decorrelation. The reduction of back-transfer is accompanied
by weaker stochastic phase scattering of the zonal shear. This suggests
a picture in which back-transfer events induce stochastic scattering
of the zonal shear phase. Suppression of back-transfer increases the
shear life time $\tau_{E}$ dominantly by reducing the phase scattering
and increases the zonal Kubo number $K_{u}$. This results in more
resilient and coherent shear patterns. Near marginal stability, back-transfer
is strongly reduced, resulting in a sharp increase in shear coherence
and in the effective shearing rate $\omega_{E}/\gamma_{max}$, even
as the absolute shear $\omega_{E}$ decreases. This reflects the growing
relative importance of shear regulation near threshold. These effects
are particularly pronounced in negative triangular (NT) plasmas, which
exhibit longer-lived and more radially extended shear layers, and
hence larger r.m.s zonal shearing rates, than do positive triangularity
(PT) configurations. The shear layers are more resilient for NT than
for PT due to the reduced back-transfer events. NT exhibits a shorter
back-transfer residence time and substantially lower excess kurtosis,
reflecting sparser and weaker bursts. Consequently, the zonal Kubo
number is higher for NT than for PT. Remarkably, NT plasmas sustain
stronger zonal $E\times B$ shear despite having lower zonal kinetic
energy, whereas PT plasmas exhibit the opposite trend, demonstrating
that turbulence regulation is controlled by zonal shear rather than
by zonal flow energy. Consistent with this picture, the turbulent
heat diffusivity is lower in NT than PT well above marginality, while
the two converge near threshold. The free energy transfer analysis
thus directly links enhanced zonal shear coherence in NT to reduced
zonal back-transfer. These results are further summarized in the following
table. 
\begin{table}[h]
\begin{tabular}{|>{\raggedright}p{2cm}|>{\centering}p{2cm}|c|>{\centering}p{2cm}|}
\hline 
Metric & PT  & NT & Implications\tabularnewline
\hline 
\hline 
Back-transfer events  & Frequent/

Strong & Reduced & NT shear is more resilient\tabularnewline
\hline 
Normalized shearing rate$(\omega_{E}/\gamma_{max})$ & Low & High & Better regulation in NT\tabularnewline
\hline 
Shear and corrugations life time

($\tau_{E}$ and $\tau_{corrugations}$) & Shorter & Longer & Persistent micro-barriers in NT\tabularnewline
\hline 
Zonal Kubo number $K_{u}=\omega_{E}\tau_{E}$ & Low & High & Better efficacy of zonal shear in NT\tabularnewline
\hline 
Zonal Kinetic Energy & High  & Low & Energy$\ne$

Shear\tabularnewline
\hline 
\end{tabular}

\caption{Summary on how nonlinear free energy transfer saturates zonal flows
in NT and PT. }

\end{table}
Overall, these results suggest that nonlinear back-transfer must be
included in reduced models of drift-wave--zonal-flow turbulence.
This can be accomplished by introducing an effective nonlinear damping
parameter, calibrated using data-driven methods. 

Finally, we speculate the local magnetic shear reversal near the outboard
midplane --- detailed in Appendix (\ref{sec:LMS}) --- as the most
plausible geometric origin of reduced back-transfer in NT. The deeper
and broader negative $\tilde{s}$-well in NT, previously shown to
stabilize linear ITG modes\citep{Singh_2025}, plausibly extends its
stabilizing influence to zonal modes by reducing their susceptibility
to back-transfer by altering the population and strength of triads
driving forward transfer and backward transfer along the field line
--- a question demanding substantial analysis is deferred to future
work. We note that the novel physics discovered here in the adiabatic
electron limit are expected to survive qualitatively with fully kinetic
electrons, with quantitative modifications to back-transfer statistics,
phase diffusion and the zonal shearing Kubo number arising from the
more complex turbulence dynamics that kinetic electrons introduce,
namely trapped electron mode (TEM) and electron temperature gradient
(ETG) driven turbulence regimes. Additionally, the present free energy
transfer diagnostic tracks the transfer of total free energy $E_{k}=S_{k}+W_{k}$,
in which the geodesic-sideband coupling cancels\citep{navarro_pop:2011}.
Separately resolving the geodesic and Reynolds-stress contributions
to back-transfer --- and quantifying how their relative weights change
with triangularity --- is an important diagnostic extension deferred
to future work. At finite $\beta$ approaching the Alfvenic regime,
magnetic field-line stochasticity can independently limit zonal flow
coherence via magnetic flutter transport, competing with the electrostatic
back-transfer channel identified here. The geometric parameters held
fixed here --- safety factor $q$, global magnetic shear $\hat{s}$,
elongation $\kappa$, and aspect ratio $a/R$ --- all influence the
zonal flow drive and damping, and may interplay with triangularity
on back-transfer. Collisionality can modulate the back-transfer events
via neoclassical zonal flow damping. Quantifying these effects are
deferred to future work. More broadly, the back-transfer--limited
zonal coherency picture developed here --- wherein intermittent back-transfer
events are linked to finite life time that governs the efficacy of
transport regulation --- provides a novel nonlinear framework that
invites analogous investigations across the many natural systems where
zonal flow coherence time plays a decisive role: from geophysical
turbulence to geodynamo convection to astrophysical disk turbulence.

\section*{Acknowledgements}

We thank Runlai Xu for helpful discussions. We acknowledge discussions
with participants in the Festival de Theory 2026. The simulations
were performed on the EXPANSE machine in San Diego Supercomputer Center.
This research was supported by U.S. DOE under Award No. $DE-FG02-04ER54738$.

\section*{Data availability}

The data that support the findings of this article are not publicly
available upon publication because it is not technically feasible
and/or the cost of preparing, depositing, and hosting the data would
be prohibitive within the terms of this research project. The data
are available from the authors upon reasonable request.


%

\appendix

\section{Coherency of corrugations\label{sec:corrg}}

\begin{figure}[h]
\includegraphics[scale=0.23]{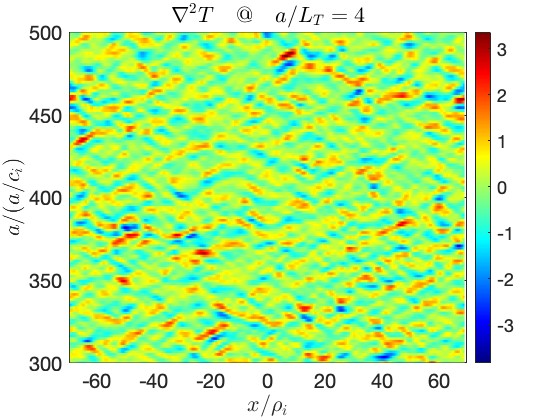}\includegraphics[scale=0.23]{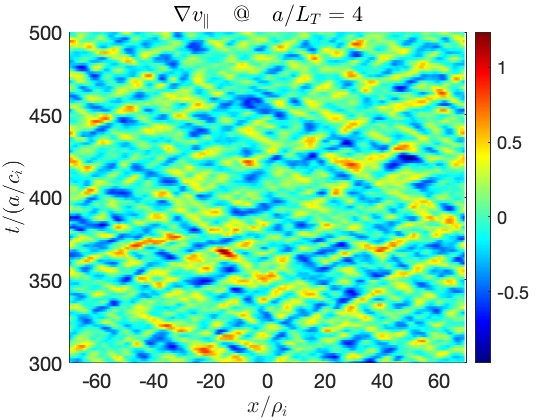}

\includegraphics[scale=0.23]{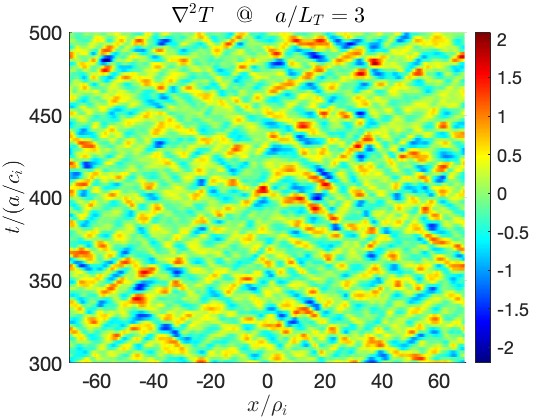}\includegraphics[scale=0.23]{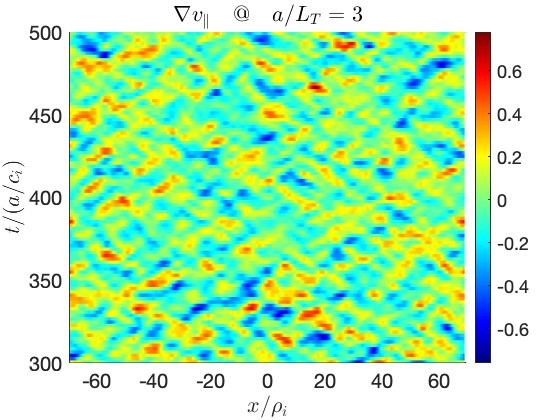}

\includegraphics[scale=0.23]{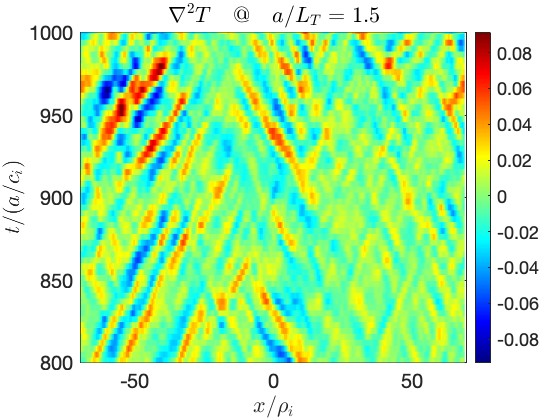}\includegraphics[scale=0.23]{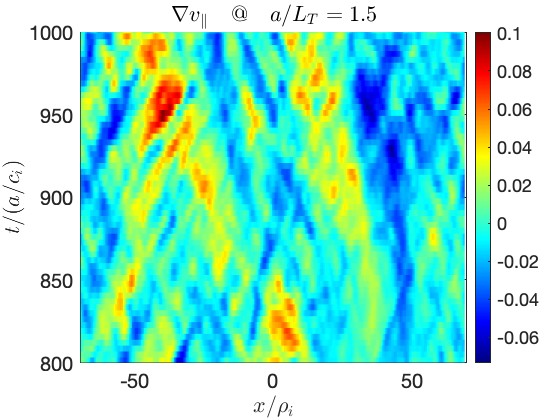}

\caption{Spatiotemporal pattern of temperature corrugations and parallel velocity
corrugations across temperature gradient for $\delta=-0.6$. Notice
the improvement in coherence as $a/L_{T}$ decreases. \label{fig:stp-1}}
\end{figure}

Corrugations participate in the heat flux avalanche dynamics and also
play important role in turbulence saturation. Corrugations enhance
eddy tilting by zonal flow shearing that manifest as enhanced k-space
diffusivity of action density due to spectral anti-correlation of
zonal potential and corrugations\citep{RS_PD_2021}. Corrugations
can also cause nonlinear growth of turbulence. The spatiotemporal
pattern of temperature and parallel velocity corrugations are shown
in the figure(\ref{fig:stp-1}).These plots reveal several notable
features. The corrugations' spatiotemporal pattern gradually becomes
more coherent with decreasing $a/L_{T}$. Near marginality, the coherence
increases drastically. The propagation speed of the corrugation fronts
also decreases as marginality is approached, as indicated by the changing
tilt of the coherent corrugation fronts. As expected, the corrugation
amplitude decreases with decreasing $a/L_{T}$, consistent with the
reduced turbulence drive.

The coherence of the corrugations is quantified next using a spatiotemporal
autocorrelation analysis. (\ref{fig:cltbt}) shows the corrugation
lifetime plotted against the zonal free-energy back-transfer, revealing
that the corrugation lifetime increases as zonal back-transfer events
are reduced. Thus, the improved coherence of corrugations is again
linked to the reduction of zonal back-transfer events as the system
approaches marginal stability from above.

\section{Gyrokinetic free energy balance\label{sec:NM}}

Multiplying the gyrokinetic Vlasov equation for the perturbed gyrocenter
distribution function by the perturbed gyrocenter distribution function
$\delta f$ and using the quasineutrality equation it is straight
forward to obtain the free energy balance equation in the wavenumber
space: 

\begin{eqnarray}
\frac{d}{dt}\left(E_{k}\right) & = & L_{T}^{-1}Q_{k}+T_{k}+D_{k}\label{eq:nlt-1}
\end{eqnarray}
Here, $E_{k}$ is the gyrokinetic free energy or generalized energy,
which is a sum of the gyrocenter entropy contribution $S_{k}$ and
electrostatic energy contribution $W_{k}$ i.e., $E_{k}=S_{k}+W_{k}$
with 
\begin{eqnarray}
S_{k} & = & \left\langle \int d^{3}v\frac{T_{0}\left|\delta f_{k}\right|^{2}}{2F_{M}}\right\rangle 
\end{eqnarray}
\begin{eqnarray}
W_{k} & = & \frac{n_{0}T_{0}}{2}\left\langle \left[1-\Gamma_{0}(b)+\left(1-\delta_{k_{y},0}\right)\tau\right]\left|\frac{e\phi_{k}}{T_{0}}\right|^{2}\right\rangle .
\end{eqnarray}
The angular bracket represents volume averaging. Notice that $E_{k}$
can also be viewed as phase-space average of the fluctuation Boltzmann
entropy expanded up to 2nd order in perturbed particle distribution
function. $E/T_{0}=-\int d^{3}xd^{3}vf^{p}\ln f^{p}\to\int d^{3}xd^{3}v\frac{\left(\delta f^{p}\right)^{2}}{2F_{M}}$,
where $\delta f^{p}$ is perturbed particle distribution function,
distinct from the gyrocenter distribution function $\delta f$. $Q_{k}$
is heat flux spectra:
\begin{eqnarray}
Q_{k} & = & \Re\left\langle ik_{y}\int d^{3}v\psi_{k}^{\star}\delta f_{k}\frac{\left(mv_{\parallel}^{2}/2+\mu B\right)}{T_{0}}\right\rangle .
\end{eqnarray}
$D_{k}$ is dissipation rate spectra: 
\begin{eqnarray}
D_{k} & = & \Re\left\langle \int d^{3}vC(\delta f_{k})\frac{T_{0}h_{k}^{\star}}{F_{M}}\right\rangle ,
\end{eqnarray}
where 
\begin{eqnarray}
h_{k} & = & \delta f_{k}+\frac{e\psi_{k}}{T_{0}}F_{M}
\end{eqnarray}
,with gyro-averaged potential $\psi_{k}=J_{0}\phi_{k}$, is the non-adiabatic
part of the perturbed distribution function. $\tau=T_{0i}/T_{0e}$
is temperature ratio. $D_{k}$ is the dissipation rate due to \textbf{numerical
hyper-diffusion in real and velocity space added for numerical stability}.
Notice that there is \textbf{no collisional dissipation} included.
$T_{k}$ is rate of the free energy transfer to mode $\vec{k}$: 
\begin{eqnarray}
T_{\vec{k}} & = & \sum_{\vec{p},\vec{q}}\mathcal{T}_{\vec{k}}^{\vec{p},\vec{q}}\label{eq:etf}
\end{eqnarray}
 defined by means of the momentum conserving triad transfer function
$\mathcal{T}_{\vec{k}}^{\vec{p},\vec{q}}$ : 
\begin{multline}
\mathcal{T}_{\vec{k}}^{\vec{p},\vec{q}}=\left\langle \frac{c}{B}\vec{b}\cdot\left(\vec{p}\times\vec{q}\right)\int d^{3}v\frac{T_{0}}{2F_{M}}\Re\left[\psi_{p}h_{q}h_{k}-\psi_{q}h_{p}h_{k}\right]\right\rangle \\
\times\delta_{\vec{k}+\vec{p}+\vec{q}=0}\label{eq:TF}
\end{multline}
Notice that $T_{k}$ results from the $E\times B$ convective nonlinearity.
This term enables free energy transfer among the modes while conserving
the total free energy via $\sum_{\vec{k}}T_{\vec{k}}=0$. Equation(\ref{eq:nlt-1})
essentially tells that the free energy at a scale $\vec{k}$ changes
due to local production $Q_{k}$, dissipation $D_{k}$ and transfer
$T_{k}$ to and from the wave number $\vec{k}$. Entropy transfer
among a group of modes can be estimated by a means of the subspace
entropy transfer analysis. For this purpose, we divide the entire
$k$-space into desired number of subspaces, maintaining the global
conservation properties. Here, we consider two subspaces:
\begin{enumerate}
\item zonal subspace with $k_{y}=0$
\item Turbulence with $k_{y}\ne0$,
\end{enumerate}
and write the entropy balance equation for the two subspaces. After
integration of equation(\ref{eq:nlt-1}) over the non-zonal modes
$k_{y}\ne0$, the free energy balance relation for the turbulent subspace
becomes: 

\begin{eqnarray}
\frac{d}{dt}\left(E_{turb}\right) & = & L_{T}^{-1}Q-\mathcal{T}_{zonal}+D_{turb},\label{eq:tee}
\end{eqnarray}
where $E_{turb}=S_{turb}+W_{turb}=\sum_{\vec{k},k_{y}\ne0}(S_{\vec{k}}+W_{\vec{k}})$.
Similarly, after integration over the zonal modes $k_{y}=0$, the
free energy balance equation for the zonal subspace becomes

\begin{eqnarray}
\frac{d}{dt}\left(E_{zonal}\right) & = & \mathcal{T}_{zonal}+D_{zonal},\label{eq:zee}
\end{eqnarray}
where $E_{zonal}=S_{zonal}+W_{zonal}=\sum_{k_{x},k_{y}=0}(S_{\vec{k}}+W_{\vec{k}})$.
Here, $\mathcal{T}_{zonal}$ represents rate of transfer of zonal
free energy due to transfer of entropy to and from the zonal mode.
Note that \emph{the zonal free energy transfer rate 
\begin{eqnarray}
\mathcal{T}_{zonal} & = & \sum_{k_{x}}T_{k_{x}}=\sum_{k_{x}}\sum_{\vec{p},\vec{q}}\mathcal{T}_{\vec{k}}^{\vec{p},\vec{q}}\label{eq:TZ}
\end{eqnarray}
is a gyro-kinetic extension of total production rate of the zonal
flow energy due Reynolds power plus corrugations energy production
due to heat flux, particle and parallel momentum flux etc. }$\mathcal{T}_{zonal}>0$
means free energy transfer from turbulence to zonal mode and $\mathcal{T}_{zonal}<0$
means back-transfer of free energy from zonal mode to turbulence.
$W_{turb}$ and $W_{zonal}$ are defined explicitly as follows:
\begin{eqnarray}
W_{turb} & = & \sum_{k_{x},k_{y}\ne0}\frac{n_{0}T_{0}}{2}\left\langle \left[1-\Gamma_{0}(b)+\tau\right]\left|\frac{e\phi_{k}}{T}\right|^{2}\right\rangle \\
 &  & =FKE+FPE
\end{eqnarray}
and 
\begin{eqnarray}
W_{zonal} & = & \sum_{k_{x},k_{y}=0}\frac{n_{0}T_{0}}{2}\left\langle \left[1-\Gamma_{0}(b)\right]\left|\frac{e\phi_{k}}{T}\right|^{2}\right\rangle \\
 &  & =ZKE
\end{eqnarray}
Notice that $W_{zonal}$ is zonal kinetic energy (ZKE) and $W_{turb}=FKE+FPE$
is sum of fluctuation kinetic energy (FKE) and fluctuation potential
energy(FPE). Integrating equation(\ref{eq:nlt-1}) over the entire
wave number space or equivalently summing the subspace entropy balance
equations(\ref{eq:tee}) and (\ref{eq:zee}) yields the global free
energy balance equation: 
\begin{eqnarray}
\frac{d}{dt}\left(E\right) & = & L_{T}^{-1}Q+D,\label{eq:geb}
\end{eqnarray}
where $(Q,D)=\sum_{k}(Q_{k},D_{k})$. Each term of the free energy
balance equations (\ref{eq:nlt-1}),(\ref{eq:tee}),(\ref{eq:zee})
and (\ref{eq:geb}) are studied with respect to variations in drive
$a/L_{T}$ and triangularity $\delta$. \\
It is worth noting the relationship between $\mathcal{T}_{zonal}$
and the geodesic-sideband transfer mechanism identified in fluid-model
studies of turbulence--flow energy exchange\citep{Scott_pla:2003,Naulin_pop:2005}.
In the fluid picture, the geodesic curvature coupling generates a
linear transfer term between the $m=0$ zonal flow and the $m=\pm1$
poloidal sidebands in the zonal kinetic energy equation. In the gyrokinetic
free energy formulation adopted here, this linear coupling does not
appear explicitly: when the dynamical equations for the electrostatic
energy $W_{k}$ and gyrokinetic entropy $S_{k}$are combined into
the total free energy $E_{k}=S_{\vec{k}}+W_{\vec{k}}$ all the linear
transfer terms including the geodesic-transfer terms cancel exactly,
leaving $\mathcal{T}_{zonal}$ arising solely from the $E\times B$
convective nonlinearity\citep{navarro_pop:2011}. Accordingly, $\mathcal{T}_{zonal}$
as defined in equation(\ref{eq:TZ}) does not separately resolve the
geodesic-sideband contribution, though the gyrokinetic simulations
themselves contain this effect self-consistently in the zonal shearing
field. Isolating the geodesic contribution within the electrostatic
energy transfer budget is left for future work.

\section{Local magnetic shear\label{sec:LMS}}

\begin{figure}[H]
\includegraphics[scale=0.23]{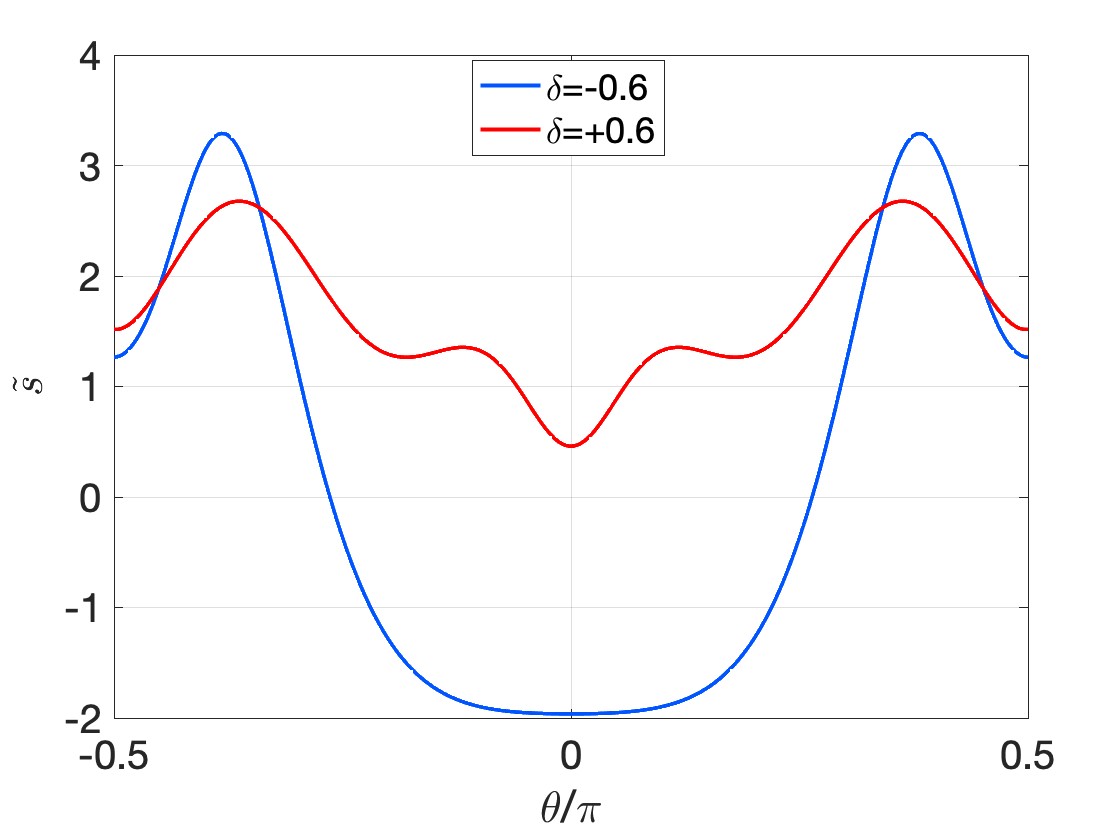}\caption{The local magnetic shear $\tilde{s}$ reverses for PT$\to$NT. \label{fig:lms}}
\end{figure}

The local magnetic shear $\tilde{s}$ is defined as the poloidal-angle-resolved
rate of change of the field line pitch: $\tilde{s}=r\nu^{\prime}/\nu$,
where the the local safety factor $\nu$ is the ratio of the toroidal
to poloidal magnetic field, and the prime denotes a radial derivative.
Unlike the global magnetic shear $\hat{s}=(r/q)(dq/dr)$, which characterizes
the average field line divergence across flux surfaces, $\tilde{s}(\theta)$
encodes how rapidly neighboring field lines separate at each poloidal
location $\theta$ along the flux tube, and is therefore the quantity
that directly governs local mode stability and eddy geometry. The
structure of the local magnetic shear $\tilde{s}(\theta)$ about the
outboard midplane is shown in figure(\ref{fig:lms}). Clearly the
$\tilde{s}$ becomes negative when triangularity changes from $\delta=+0.6$
to $\delta=-0.6$. Moreover, the negative shear well is deeper and
broader for NT than for PT. This reduces the eigenmode averaged magnetic
drift frequency for NT, thus reducing the ITG drive\citep{Singh_2025}.
Also, a locally reversed magnetic shear twists eddies in a short distance
along the field line to point along the good curvature\citep{Antonsen_1996,Gaur_2023,Beeke_2021}.
Thus, negative local shear is stabilizing. 

\section{Parameters}

We consider the following shaping parameters $a/R=1/3$, flux-tube
location $r=a/2$, safety factor $q=2$, magnetic shear $\hat{s}=\frac{r}{q}\frac{dq}{dr}=1$,
triangularity $\delta$ varied in the range $-0.6\le\delta\le+0.6$,
elongation $\kappa=1$, triangularity gradient $S_{\delta}=\frac{r}{\sqrt{1-\delta^{2}}}\frac{\partial\delta}{\partial r}=\frac{\delta}{\sqrt{1-\delta^{2}}}$,
elongation gradient $S_{k}=\frac{r}{\kappa}\frac{\partial\kappa}{\partial r}=\frac{\kappa-1}{\kappa}$,
squareness $\zeta=0$, squareness gradient $S_{\zeta}=\frac{r}{\zeta}\frac{\partial\zeta}{\partial r}=0$,
elevation $Z_{0}=0$, elevation gradient $S_{Z}=\frac{r}{Z_{0}}\frac{\partial Z_{0}}{\partial r}=0$,
MHD alpha parameter $\alpha_{MHD}=0$, Shafranov shift gradient $R_{0}^{\prime}=0$.
The ion temperature gradient varied in the range $a/L_{T_{i}}=[1.5-4]$,
and the plasma density gradient is $a/L_{n}=1$. Electrons response
is considered adiabatic with $T_{e}=T_{i}$. For nonlinear simulations,
following grid resolutions are used: $n_{x}=257$, $n_{k_{y}}=48$,
$n_{z}=64$, $n_{v_{\parallel}}=48$, $n_{\mu}=8$. The simulation
box size in radial direction is $L_{x}=[120-140]\rho_{i}$ , along
$v_{\parallel}$ direction is $L_{v_{\parallel}}=3c_{i}$, along $\mu$
direction is $L_{\mu}=9\mu B_{0}/T_{i}$, where $c_{i}=\sqrt{2T_{i}/m_{i}}$
is the thermal speed of ion of mass $m_{i}$ and temperature $T_{i}$
. The smallest $k_{y}$ resolution $k_{y,min}\rho_{i}=0.05$ and the
numerical hyper-diffusion coefficients along $z$ direction is $\varepsilon_{z}=2$
and that along $v_{\parallel}$ direction is $\varepsilon_{v}=0.2$.
The hyper-diffusion coefficients are chosen to be consistent with
those used in previous studies\citep{Duff_etal_2022,Merlo_2023,Singh_2025}
employing the same set of other parameters. The reference values in
SI units are the default values used in GENE i.e, $m_{i}=3.33\times10^{-27}kg$,
$T_{i}=400eV$, $B_{0}=2T$, $a=0.8m$, $n=1.65\times10^{19}m^{-3}$. 
\end{document}